\documentstyle[a4,12pt,axodraw]{article}
\def\be{\begin{equation}}
\def\ee{\end{equation}}
\def\bq{\begin{eqnarray}}
\def\eq{\end{eqnarray}}
\begin{document}
\thispagestyle{empty}
\setcounter{page}{0}
\begin{flushright}
MPI-PhT/94-26\\
LMU 05/94\\
May 1994
\end{flushright}
\vspace*{\fill}
\begin{center}
{\Large\bf QCD Calculation of $B$ Meson Form Factors and Exclusive
Nonleptonic Decays}$^*$\\
\vspace{2em}
\large
A. Khodjamirian$^{a,b,1}$ and R. R\"uckl$^{a,c,2}$\\
\vspace{2em}
{$^a$ \small Sektion Physik der Universit\"at
M\"unchen, D-80333 M\"unchen, Germany }\\
{$^b$ \small Yerevan Physics Institute, 375036 Yerevan, Armenia } \\
{$^c$ \small Max-Planck-Institut f\"ur Physik, Werner-Heisenberg-Institut,
D-80805 M\"unchen, Germany}\\

\end{center}
\vspace*{\fill}

\begin{abstract}
The method of
QCD sum rules has proven to be particularly useful in heavy quark
physics, where a small distance scale is provided
by the inverse heavy quark mass. We present two new examples for
the application of this method to B-physics:
a calculation of the heavy-to-light form factors $B \rightarrow
\pi,K$ and an estimate of the weak decay amplitude for
$B \rightarrow J/\psi K $
beyond the factorization approximation.
\end{abstract}

\vspace*{\fill}

\begin{flushleft}
\noindent$^1$ Alexander von Humboldt Fellow\\
\noindent$^2$ supported in part by the German Federal Ministry for Research and
Technology (BMFT) under contract No. 05 6MU93P\\
\noindent$^*$ {\it presented by A. Khodjamirian at the Workshop
"Continuous Advances in QCD", TPI, University of Minnesota,
Minneapolis, February 18-20, 1994 }
\baselineskip=16pt
\end{flushleft}

\newpage
\section{Introduction}

Introduced fifteen years ago, the method of QCD sum rules \cite{SVZ}
has become one of the most useful and reliable calculational
schemes to derive hadronic properties from first
principles with a minimal number of additional phenomenological
assumptions.
Originally, focusing on the operator product expansion (OPE)
of two-point correlation functions one aimed at the
determination of masses and coupling constants of ground
states. The framework has then been extended and also applied
to other dynamical characteristics of hadrons such as
form factors and decay amplitudes. These extensions are based on
OPE of three- and four-point  correlation functions.
Nevertheless, no new input parameters appear
which is one of the most attractive features of this approach.
All essential parameters such as vacuum
condensates, coupling constants and thresholds of higher states
are fixed by corresponding two-point sum rules.

On the other hand, straightforward application of QCD sum rules in
their original version to many-point correlation functions runs
into a number of technical problems.
First of all, the correlation functions are to be calculated in
the deep Euclidean region. The result is then continued to
physical momenta with the help of dispersion relations.
In the presence of many channels, this continuation is not at all
trivial. In some cases like for on mass-shell hadronic vertices
and amplitudes one must rely on extrapolations which
are plausible but not rigorous. Furthermore, the OPE for
n-point correlation functions is usually represented by
Feynman-type diagrams which are used to calculate the Wilson
coefficients in perturbation theory. These diagrams grow rapidly
in number with n and become rather complicated. Finally, in order
to include higher resonances and continua of states into the
sum rules, one often has to resort to models based on the
concept of quark-hadron duality. These models usually work
well for two-point correlation functions, but  may be
too crude already in the case of double dispersion integrals.

For these and other reasons, it is very important to develop
alternative, more
economical methods which avoid these problems. Clearly, starting
from vacuum correlation functions, applying OPE and deriving
sum rules in terms of local vacuum condensates
is only one way to account for nonperturbative
quark-gluon dynamics. Other approaches leading to different
versions of QCD sum rules include the external field  method
\cite{IofSmil}-\cite{BelKog} and
induced vacuum condensates, the concept of nonlocal condensates
\cite{MikhRad}, and the combination with light-cone wave functions
\cite{ChernZhit}-\cite{BrHalp}. Each of these versions has
advantages and disadvantages. The comparison of the results may
shed some light on the accuracy and reliability of the different
approaches.

In this report we want to illustrate some of the current trends
in the development of the QCD sum rule method by presenting two
recent calculations relevant for B-physics. In section 2 we discuss
the calculation of the
$B \rightarrow \pi$ and $B \rightarrow K $ form factors using
an expansion in terms of pion  and kaon wave functions  on the
light-cone with increasing twist, and compare the results with
the predictions of conventional QCD sum rules. Then, in section 3,
we study the problem how to calculate the
amplitude of the prominent weak decay $ B \rightarrow J/\psi K $
beyond the usual factorization approximation. In other words, we
undertake the difficult task to estimate the nonfactorizable
contributions to the decay amplitude originating from quark-gluon
interactions at scales between the heavy b-quark mass and
the hadronization scale. Following an idea suggested in ref. [11]
for $D$ meson nonleptonic decays, we try to extract this amplitude
from a suitable four-point correlation function treated within the
framework of conventional QCD sum rules.

\section{QCD sum rules on the light-cone for the
$B \rightarrow \pi, K$ form factors}

The standard way to derive QCD sum rules for the transition
amplitude $ <A|j|B> $ between given ground state hadrons $A$ and
$B$ starts from the vacuum correlation function of three
currents, $ <0|T \{j_A(x) j(0) j_B(y)\}|0>$, where $j_A$ ($j_B$)
is chosen such as to create the hadron $A$ ($B$) from the
vacuum and $T$ denotes the time-ordered product. A double
dispersion relation in $p_A^2$ and $p_B^2$, $p_A$ ($p_B$) being
the four-momentum in the $A$ ($B$)-channel,  is then used to
relate this correlation function calculated in the Euclidean region
with its imaginary part containing the desired transition
amplitude.
The remainder of the calculation is essentially the same
as in the case of two-point QCD sum rules: derivation of OPE in
terms of local quark-gluon condensates,
subtraction of excited state and continuum contributions and
Borel transformation.
Since the quark-gluon condensates are universal, they can
be taken from any QCD sum rule analysis. Furthermore,
the coupling constants
$f_A \sim <0|j_A|A>$ and $f_B \sim <0|j_B|B>$ as well as the
thresholds of higher states in the $A$ and $B$ channels are fixed
by the two-point sum rules for the correlation functions
$<0 \mid j_A j_A \mid 0>$ and $<0 \mid j_B j_B \mid 0>$,
respectively. Hence, as already pointed out, one has essentially
no new parameters.
This procedure has been applied to a variety of problems such as
the pion form factor \cite{IofSmil1,NestRad}, radiative charmonium
transition rates \cite{Kh1}-\cite{RRY}, and semileptonic form
factors of heavy mesons \cite{AlEl}-\cite{Nar}.

Here we demonstrate an alternative method which may be
used in cases where one of the hadrons, $A$ or $B$,  is a light
meson. An important example is provided by the $B \rightarrow \pi$
form factors
$f^{\pm}_{\pi}$ which determine the matrix element
\be
<\pi|\bar{u} \gamma_\mu b |B>= 2f_\pi^+(p^2)q_\mu +
[f_\pi^+(p^2)+f_\pi^-(p^2)]p_\mu
\label{form}
\ee
with $p+q$, $q$ and $p$ being the $B$ and $\pi$ momenta and the
momentum transfer, respectively. In the following,
we concentrate on the phenomenologically more interesting form
factor $f^+_\pi$.
The corresponding $B \rightarrow K$ form factor $f^+_K $ can be
treated in parallel by obvious formal replacements. Numerical
results will be shown for both  $f^+_\pi$ and  $f^+_K $.

Instead of investigating the vacuum averaged correlation of the
$b \rightarrow u$ transition current with two other currents
carrying the  $B$ and $\pi$ quantum numbers, we consider the matrix
element
\be
F_\mu (p,q)=
i \int d^4xe^{ipx}<\pi(q)\mid T\{\bar{u}(x)\gamma_\mu b(x),
\bar{b}(0)i\gamma_5 d(0)\}\mid 0>
\label{1a}
\ee
$$
= F((p+q)^2,p^2) q_\mu + \tilde{F}((p+q)^2,p^2) p_\mu
$$
between the vacuum and an on-shell pion state.
This object is represented diagrammatically in
Fig. 1. The pion momentum squared, $q^2 = m_\pi^2$, vanishes in the
chiral limit adopted throughout this discussion. Moreover,
the light $u$ and $d$ quarks forming the pion eventually propagate
to large distances.
 In contrast, the b-quark still propagates far off-shell
provided that $(p+q)^2$ is taken sufficiently large and negative,
and the time-like momentum transfer squared  $p^2$ is far from the
kinematical limit, $p^2 = m_B^2 $.

Formally, contracting the $b$-quark fields in (\ref{1a}) and keeping
only the lowest order term, i.e. the free $b$-quark propagator,
yields
\be
F ((p+q)^2,p^2)=
i \int d^4x \int \frac{d^4P}{(2\pi)^4}e^{i(p-P)x}
\sum_a \frac{\phi_a(x^2,q \cdot x)}{P^2-m_b^2}~,
\label{F}
\ee
where
\be
\phi_a(x^2, q\cdot x) =
<\pi(q)\mid \bar{u}(x)\Gamma_a d(0)\mid 0>~,
\label{phi}
\ee
$\Gamma_a$ denoting certain combinations of Dirac matrices. This
approximation corresponds to Fig. 1a.
The high virtuality of the $b$-quark propagating between the points
$x$ and $0$ guarantees that the $u$ and $d$ quarks are
emitted at almost light-like distances. In that case, it is
justified to keep only the first few terms in the expansion
of the matrix elements (\ref{phi}) around $x^2=0$, that is near
the light-cone:
\be
\phi_a(x^2, q \cdot x) = \sum_n \phi_a^n(q \cdot x) (x^2)^n
\label{phi1}~.
\ee
Logarithms in $x^2$ which may also appear in (\ref{phi1}) are
disregarded for simplicity. These terms can
be consistently treated by means of QCD perturbation theory.
They give rise to normalization scale dependence.
Because of translation invariance, the coefficients $\phi_a^n$
must have the form
\be
\phi_a^n (q\cdot x) \sim \int_0^1 du\varphi_a^n(u)exp(iuq\cdot x)~.
\label{varphi}
\ee
Inserting (\ref{phi1}) and (\ref{varphi}) into (\ref{F}) and
integrating over $x$ and $P$, one obtains, schematically,
\be
F((p+q)^2, p^2) \sim
\sum_a \sum_{n } \int^1_0 du \frac{\varphi^n_a(u)}
{[m_b^2-(p+qu)^2]^{2n}}~.
\label{F1}
\ee
It is thus possible to
calculate the invariant function $F$ with reasonable accuracy in the
kinematical region of highly virtual b-quarks provided one
knows the distribution functions $\varphi_a^n(u)$
at least for low values of $n$.
These distributions contain essential information about
the dynamics at large distances and play a similar
role  as the vacuum condensates in the conventional
OPE of vacuum correlation functions.

As it turns out, the distributions
$\varphi_a^n(u)$ are nothing but light-cone wave functions of the
pion the twist of which is growing with $n$. These wave functions
were introduced long ago in the context of hard exclusive
processes \cite{ChernZhit1}-\cite{FarrJacks}.
The leading twist 2 wave function is defined by
\be
< \pi(q) \mid \bar{u}(x) \gamma_\mu \gamma_5 P exp
\{ i \int ^1_0d\alpha~ x_\mu A^\mu (\alpha x) \} d(0) \mid 0 >=
-if_\pi q_\mu \int ^1_0 du e^{iuq \cdot x} \varphi_\pi(u)~,
\label{tw2}
\ee
where the exponential factor involving the gluon field is necessary
for gauge invariance.
The asymptotic form of $ \phi_\pi(u)$ can be obtained from
perturbative considerations and is well known :
\be
\varphi_\pi(u)=6u(1-u).
\label{as}
\ee
Over the years a great deal has been learned about these wave
functions \cite{CrSt}-\cite{BrFil}.
All important twist two, three and four wave functions have been
identified and their asymptotic form has been determined. Besides
two-particle wave functions reflecting the quark-antiquark
structure of mesons, also three-particle wave functions
associated with the quark-antiquark-gluon component have been
studied. Very important for practical applications,
nonasymptotic corrections have been estimated considering
expansions in orthogonal polynomials and renormalizing the
coefficients of this expansions in QCD perturbation theory.
Furthermore, conventional two-point QCD sum rules have been used
\cite{ChernZhit2} to fix  the first few moments of the low
twist wave functions. Based on these estimates and improvements,
various models for the nonasymptotic form of
these functions have been suggested
( see, for example, refs. [10,28,32]).

The approach outlined above together with the accumulated knowledge
of pion and kaon wave functions has been employed in
ref. [6] to calculate the $ B \rightarrow \pi$ form
factor at zero momentum transfer, in ref. [7]  to get the
$ D\rightarrow \pi$ form factor, and in ref. [8] to derive
the $B\rightarrow \pi$ and $B \rightarrow K$ form factors including
the momentum dependence.

In our calculation of $f^+_\pi$ and  $f^+_K$ we have included
quark-antiquark wave functions up to twist four. In addition, we
have also estimated the first-order correction to the free
$b$-quark propagation shown in Fig. 1b which involves
quark-antiquark-gluon wave functions. On the other hand, the
perturbative $O(\alpha_s)$ corrections indicated in Figs. 1c and
1d  have not been evaluated directly, but have been taken
into account indirectly as explained below. The direct
calculation of these corrections is on its way.

In order to extract the desired form factor from the resulting
invariant function $F((p+q)^2,p^2)$ sketched in (\ref{F1})
we employ a QCD sum rule with respect to the $B$-meson channel.
Writing  a dispersion relation in $(p+q)^2$, we approximate the
hadronic spectral function in the $B$-channel by the pole
contribution of the $B$ meson and a continuum contribution.
In accordance with quark-hadron duality, the latter is
identified with the spectral function derived from the QCD
representation (\ref{F1}) above the threshold $(p+q)^2=s_0$.
Formally, subtraction of the continuum then amounts to simply
changing the lower integration boundary in (7) from 0 to
$\Delta = (m_b^2-p^2)/(s_0-p^2) $. After Borel transformation
one arrives at a sum rule for the product $f_\pi^+f_B$ where
$f_B$ is the $B$ meson decay constant.
It is important to note that the numerical values to be substituted
for $m_b$, $f_B$ and the threshold $s_0$ are interrelated by the
QCD sum rule for the correlation
function $<0 \mid T \{ \bar{b}(x)\gamma_5 u(x),
\bar{u}(0)\gamma_5 b(0)\} \mid 0>$.
We stress that the latter sum rule should be used without
$O(\alpha_s)$ corrections in order to be consistent with the
neglect of these corrections in the sum rule for $f_\pi^+f_B$
(see also ref. [20]).

The final expression of the form factor  $f_\pi^+$ is given by
$$
f^+_\pi( p^2)= \frac{f_\pi m_b^2}{2f_Bm_B^2}\int_
\Delta^1\frac{du}{u}
exp[\frac{m_B^2}{M^2}-\frac{m_b^2-p^2(1-u)}{uM^2}]
$$
\be
[ \varphi_\pi(u) + \frac{\mu}{m_b}u\varphi_{p}(u)
+ \frac{\mu}{6m_b}\varphi_
{\sigma }(u)(2 + \frac{m_b^2+p^2}{uM^2}) ]~,
\label{20a}
\ee
where $M^2$ is the Borel parameter and $\mu = m_\pi^2/(m_u+m_d)$.
While in the concrete calculation of ref. [8] the
twist 2 wave function $\varphi_\pi$ is corrected for
nonasymptotic effects \cite{ChernZhit2},\cite{BrFil},
the twist 3 wave functions
$\varphi_p$ and $\varphi_\sigma$ are taken in their asymptotic
form. Numerically, the higher twist contributions turn out to be
more important than the nonasymptotic corrections to the leading
twist wave function. For brevity, in (\ref{20a}) we have omitted
the contributions from the twist four quark-antiquark wave functions
and the term involving the twist three quark-antiquark-gluon wave
function. These contributions are quantitatively negligible.
Further details can be found in ref. [8].

Numerical results for $f^+_\pi( p^2)$ are plotted in Fig. 2a for two
values of the Borel parameter $M$ characterizing the fiducial
range of the method. As can be seen, the predictions are quite
stable under variation of $M$ as long as $p^2$ is not getting too
close to $m_B^2$. For comparison, we also show results
obtained by other methods. Furthermore, it is interesting to note
that  $f^+_\pi( p^2)$ is not very sensitive to the precise shape
of the leading twist wave function $\varphi_\pi(u)$  at least
at $p^2 \leq 10 GeV^2$. We have checked this by replacing the
nonasymptotic two-humped wave function
\cite{ChernZhit2} used for Fig. 2a by the simple asymptotic wave
function given in (\ref{as}).

The $B \rightarrow K $ form factor $f^+_K$ was calculated
analogous to  $f^+_\pi$ using the leading twist wave function
$\varphi_K$ from ref. [28]. The higher twist wave
functions are left unchanged.
The numerical result is presented  in Fig. 2b.
Although $f^+_K$ is not accessible directly
in semileptonic $B$ decays, it determines the factorizable part of
the nonleptonic decay amplitude for $B \rightarrow J/\psi K$ and
is therefore phenomenologically very important. Later, in section
3 we shall make use of the value
\be
f^+_K( m_{\psi}^2)= 0.55 \pm 0.05 ~.
\label{fpsi}
\ee

In conclusion, we emphasize that light-cone sum rules such as the
ones exemplified in this section represent a well defined
alternative to the conventional QCD sum rule method. In this
variant, the nonperturbative aspects are described by a set of
wave functions on the light-cone with varying twist and quark-gluon
multiplicity.
These universal functions can be studied in a variety of processes
involving the $\pi$ and $K$ meson, or other light mesons.
The most important advantage of the light-cone sum rules is the
possibility to take hadrons on mass-shell from the
very beginning.
One can thus avoid the notorious model-dependence of extrapolations
from Euclidean to physical momenta in light channels.
Furthermore, in many cases the light-cone approach
is technically much easier than the conventional
QCD sum rule technique. Finally, the light-cone method is rather
versatile. It can also profitably be employed to calculate
heavy-to-light form factors such as $B \rightarrow \rho$ and
$B \rightarrow K^*$, amplitudes of rare decays such as
$B \rightarrow K^* \gamma$ \cite{ABS}, and hadronic couplings such
as $B^*B \pi$ \cite{BBKR}. Just in passing we mention that the
$B^*B \pi$ coupling can be extracted from the correlation function
(2) by performing a second Borel transformation  in $p^2$.
Needless to say, light-cone sum rule are equally useful in
calculating properties of $D$ mesons and also
of light hadrons \cite{BrHalp},\cite{BrFil}.

The main problem to be solved if one wants to fully exploit the
light-cone approach is a reliable determination of the
nonasymptotic effects in the wave functions. In this respect,
measurements of hadronic form factors, couplings etc. can
provide important information.
A second, mainly technical problem, concerns higher order
perturbative corrections, such as those shown in Figs. 1c and 1d,
and the possible occurrence of large logarithms.

\section{ Weak decay amplitude for $B \rightarrow J/\psi K $
beyond factorization}

Since the earliest estimates \cite{EGN}$^-$\cite{CabMai},
nonleptonic two-body decays of heavy mesons are usually calculated
by splitting the appropriate matrix element of the
weak Hamiltonian into a product of a semileptonic transition
form factor and a decay constant of one of the final mesons.
On the theoretical side, one still lacks a strict proof of this
recipe. Empirically, with the accumulating data on $D$ decays, it
soon became clear that naive factorization fails \cite{Ruckl}.
In order to
achieve agreement with experiment it is necessary to let the
coefficients of the OPE of the weak Hamiltonian deviate from their
values predicted in short-distance QCD. Phenomenologically
\cite{BSW}, the coefficients are reinterpreted and treated as free
parameters to be determined from experiment. For $D$ decays, it has
been shown \cite{BGR} that the fitted values can be reconciled
with the short-distance expectations if
the matrix elements of the local operators appearing in the OPE
are expanded in $1/N_c$ and if only the leading terms are kept.
This observation gave rise to the rule of discarding nonleading in
$1/N_c$ terms. Factorization combined
with this rule leads to a satisfactory description of the majority
of two-body $D$-decays.

For $B$ decays, an instructive example is provided by the
illustrious decay mode $B \rightarrow J/\psi K $.
The effective weak Hamiltonian responsible for this decay is given
by
\be
H_W= \frac{G}{\sqrt{2}}V_{cb}V^*_{cs}\{c_1O_1+c_2O_2\}~,
\label{H}
\ee
\be
O_1=(\bar{s}\Gamma^\rho c)(\bar{c}\Gamma_\rho b),\
O_2=(\bar{c}\Gamma^\rho c)(\bar{s}\Gamma_\rho b)
\label{o}
\ee
where $G$ is the Fermi constant , $V_{cb}$ and $V_{cs}$ are the
relevant CKM matrix elements and
$\Gamma_\rho = \gamma_\rho(1-\gamma_5)$. The coefficients
$c_{1}(\mu)$ and $c_2(\mu)$ incorporate the short-distance
effects arising from the renormalization of $H_W$ from $\mu=m_W$
to $\mu=O(m_b)$. These coefficients are known to next-to-leading
order (NLO) \cite{AGMP},\cite{BurW}. While $O_2$ already
possesses the appropriate flavour structure, this is the case for
$O_1$ only after Fierz transformation:
\be
O_1= \frac13O_2+2\tilde{O}_2 ~,
\label{Fz}
\ee
where
\be
\tilde{O_2}=(\bar{c}\Gamma^\rho
\frac{\lambda^a}{2}c)(\bar{s}\Gamma_\rho
\frac{\lambda^a}{2}b)\}
\label{Otilde}~.
\ee

Assuming factorization, the matrix element of $H_W$ for
$B \rightarrow J/\psi K$ simplifies to
$$
<J/\psi K\mid (c_2+\frac{c_1}3)O_2 +2c_1\tilde{O_2} \mid B>
$$
\be
=(c_2+\frac{c_1}3)<J/\psi\mid \overline{c}\Gamma^\rho c\mid 0>
<K\mid \overline{s}\Gamma_\rho b \mid B>~.
\label{matr}
\ee
Note that in this approximation
\be
<J/\psi K \mid \tilde{O_2} \mid B> =
<J/\psi\mid \overline{c}\Gamma^\rho \frac{\lambda^a}2 c\mid 0>
<K\mid \overline{s}\Gamma_\rho \frac{\lambda^a}2 b \mid B>=0
\label{nf}
\ee
because of color conservation. Moreover, using a parametrization
of $<K\mid \overline{s}\gamma_\rho b \mid B>$  analogous to
(\ref{form}) and
$<J/\psi\mid \overline{c}\gamma_\rho c\mid 0>= m_\psi f_\psi
\epsilon_\rho $ one readily obtains the following expression
for the decay amplitude:
\be
A( B \rightarrow J/\psi K ) =
\sqrt{2}GV_{cb}V^*_{cs}(c_2+\frac{c_1}3)f_{\psi}
f^+_K(p^2=m_{\psi}^2)m_{\psi}
(\epsilon \cdot q)~.
\label{decay}
\ee
The corresponding branching ratio is given by
$$
BR( B \rightarrow J/\psi K ) = \frac{G^2}{32 \pi }
|V_{cb}V_{cs}^*|^2
(c_2 + \frac{c_1}{3})^2f_\psi^2f^{+2}_K(p^2=m_{\psi}^2)
$$
\be
\times m_B^3(1-\frac{m_{\psi}^2}{m_B^2})^3\tau_B
\label{BR}
\ee
where $\tau_B $ is the $B$ meson lifetime.

Taking for the $B\rightarrow K$ form factor the value
(\ref{fpsi}) from ref. [8] which agrees with other calculations,
and using $f_{\psi}=409~MeV$,
$V_{cb}= 0.04 $, and $\tau_B= (1.489 \pm 0.038) ~~ps $
we obtain
\be
0.005\% \leq  BR( B\rightarrow J/\psi K) \leq 0.013\% ~.
\label{BRfact}
\ee
The range of this prediction mainly reflects the theoretical
uncertainties in the coefficients $c_1$ and $c_2$ at $\mu=m_b$,
given in ref. [42]:
\be
\{c_1 =1.115,~c_2=-0.255\} \div \{c_1 = 1.146,~c_2=-0.312 \} ~.
\label{c12}
\ee
As a matter of fact, the combination $c_2+c_1/3$ turns out
to be particularly sensitive to the precise values
of $c_1$ and $c_2$, as pointed out in ref. [43]. Despite of this
uncertainty, the branching ratio expected from naive
factorization is undoubtedly much lower than the recent
experimental results \cite{CLEO}:
$$
BR( B^- \rightarrow J/\psi K^- )= (0.11 \pm 0.015 \pm 0.009)\%
$$
\be
BR( B^0 \rightarrow J/\psi K^0 )= (0.075 \pm 0.024 \pm 0.008
)\% ~.
\label{CLEO}
\ee
On the other hand, applying the rule of discarding the nonleading
in $1/N_c$ terms, that is dropping the term $c_1/3$ in
(\ref{BR}), yields
\be
0.065\% \leq  BR( B\rightarrow J/\psi K) \leq 0.095\%
\label{BRNc}
\ee
in good agreement with (\ref{CLEO}).

Yet, from this observation one cannot conclude that the
$1/N_c$-rule generally works in $B$
decays. The CLEO analysis \cite{CLEO}, for example, shows that
factorization is consistent with the measurements provided the
short distance coefficients
$c_1+c_2/3$ and $c_2+c_1/3$ are replaced by two a priori unknown
parameters $a_1$ and $a_2$, respectively. Similarly as in $D$
decays, these coefficients are found to be universal, at least on
the basis of the two-body $B$ decays analysed so far. However,
unlike in $D$ decays the data favour a value $a_2 >0 $ which
is inconsistent with the $1/N_c$-rule implying $a_2 \simeq c_2 < 0$.
( see also the discussion in ref. [45]).

The present situation is very unsatisfactory and rather confusing.
Theoretically, there seems to be no trustworthy reason why
factorization should work. Moreover, on dynamical
grounds \cite{KLR} the universality of $a_1$ and $a_2$ is rather
surprising than expected.
Also, the empirical facts do certainly not prove
factorization. Just the contrary may be the case. For example, in
the matrix element (\ref{matr}) the factorizable term proportional
to $c_1/3$ may cancel with sizeable nonfactorizable contributions
neglected in eqs.(\ref{matr}) to (\ref{BRfact}). Effectively, this
would lead to the result (\ref{BRNc}) and explain the success of
the $1/N_c$ rule. Such a cancellation was first advocated in
ref. [38] and then shown in ref. [11] to actually
take place in two-body $D$-decays by estimating the nonfactorizable
contributions to the decay amplitudes
using QCD sum rule techniques ( for a more recent discussion see
refs. [46,47]). This cancellation may be
different in $B$ decays. It may occur in certain channels and not in
others. The nonfactorizable contribution may also happen to
overcompensate the factorizable amplitudes nonleading in $1/N_c$, and
thus change even the sign of the effective coefficient $a_2$.
Clearly, one has to explain the coefficients $a_1$ and $a_2$ before
one can claim some theoretical understanding.

Recently, we have started
an attempt in this direction \cite{KLR}. We are investigating
the problem of factorization in $B$ decays using
$B \rightarrow J/\psi K $ as a study case.
Here we briefly explain our approach and present some
preliminary results.
Following the general idea suggested in ref. [11],
we consider the four-point correlation function :
\be
\Pi_{\mu\nu} (q,p)=
\int d^4xd^4yd^4ze^{iqx+ipy}<0\mid
T\{j_{\mu5}^K(x)j_\nu^\psi(y)H_W(z)j^B_5(0)
\}\mid 0> ~,
\label{corr}
\ee
where $ j_{\mu5}^K= \bar{u}\gamma_\mu \gamma_5s $ ,
$ j_\nu^\psi= \bar{c}\gamma_\nu c $ and
$ j^B_5= \bar{b}i\gamma_5 u $
are the generating currents of the mesons involved and
$H_W$ is the effective weak Hamiltonian given in (\ref{H}).
The four-momenta assigned to the $K$, $J/\psi$ and $B$ channel
are $q$, $p$ and $P=q+p$, respectively. The momentum transfer
in the weak interaction point $z$ is zero. The bare diagram
associated with (\ref{corr}) is depicted in Fig. 3a.

The correlation function $\Pi_{\mu\nu}$ is then
calculated by means of the QCD operator product expansion.
The appropriate Euclidean region located far enough from the
physical thresholds in all three channels is
$q^2 \leq -1 GeV^2 $ , $p^2 \leq 0$ and $P^2 \leq 0 $.
For the $J/\psi$ and $B$ channels, the conditions are more relaxed
than for the $K$-channel, since the former are protected by the
large $c$ and $b$ quark masses. The light quarks are considered
massless.
In the OPE for (\ref{corr}) we include all local operators
up to dimension six. The Wilson coefficients are calculated to lowest
nonvanishing order in $\alpha_s$.

In this approximation the
nonfactorizable contributions to (\ref{corr}) only arise from the
operator $\tilde{O_2}$ given in (\ref{Otilde}). This is just the
operator the matrix element of which vanishes by factorization as
pointed out in (\ref{nf}). However, nonperturbative gluon exchange
between the loop and the triangle indicated in Fig. 4 breaks
factorization and gives rise to a finite contribution from
$\tilde{O_2}$. Fig. 4a shows one of the diagrams associated with
the $d=4$ gluon operator $G_{\mu\nu}^aG^{a\mu\nu}$. Typical
diagrams for the $d=5$ quark-gluon operator
$\bar{q}gG_{\mu\nu}^a\lambda^a\sigma^{\mu\nu}q$ and the
$d=6$ four-quark operator are depicted in Figs. 4b and 4c,
respectively. Additional $d=6$ contributions corresponding to the
diagram Fig. 4b emerge from the operators
$\bar {q} \nabla_\sigma q G_{\tau \lambda}^a $ and
$\bar {q} {\cal D}_\sigma G_{\tau \lambda}^a q $, where
$\nabla_\sigma $ and $ {\cal D}_\sigma $ denote the proper
covariant derivatives.

Of course, there are also perturbative gluon corrections to the
bare diagram, Fig. 3a, associated with the unity operator in the
OPE. These are exemplified in Fig. 3b and 3c. The analogous
corrections to
two-point functions play an essential role in the corresponding sum
rules. Therefore, they should also be included in a complete
treatment of the four-point function (\ref{corr}). Some of the
perturbative gluon corrections (Fig. 3b) are in fact included,
summed up in the coefficients $c_1(m_b)$ and $c_2(m_b)$ of $H_W$,
namely those originating at short-distances between $1/m_W$ and
$1/m_b$. However, there are other contributions (Fig. 3c) coming
from larger distances, say, between $1/m_b$
and $1/m_c$ or even $O(1GeV^{-1})$, which are definitely not included
in $H_W$. These may give rise to logarithms of the kind
$~ln(m_b/m_c)$ and induce important nonfactorizable contributions
to the decay amplitudes. A systematic inclusion of all perturbative
gluon effects requires three-loop calculations with massive quarks,
a task which is postponed to later developments. Restricting
in the first step of the analysis to the nonperturbative
interactions sketched in Fig. 4, the correlation function
(\ref{corr}) receives the following nonfactorizable contributions:
\be
\tilde{\Pi}_{\mu\nu}^{QCD}=\Pi_{\mu\nu}^{GG}
+\Pi^{\bar{q}Gq}_{\mu\nu} +
\Pi^{\bar{q} \nabla q G}_{\mu\nu}+\Pi^{\bar{q} {\cal D }
Gq}_{\mu\nu}+
\Pi^{\bar{q}q\bar{q}q}_{\mu\nu} ~,
\label{res}
\ee
where the tilde is a reminder to replace $H_W$ in (\ref{corr})
by $\tilde{O_2}$. The individual terms in (\ref{res}) have been
derived from the corresponding diagrams of Fig. 4 in the form
of Feynman integrals. Their explicit expressions can be found
in ref. [48].

With the result (\ref{res}) at hand, one can now
proceed in constructing sum rules which will allow to determine
the  matrix element (\ref{nf}) entering the
correlation function through the resonance
contribution

\be
\tilde{\Pi}_{\mu\nu}^{B\rightarrow J/\psi K } =
i\frac{<0 \mid j_{\mu5}^K \mid K><0\mid j_\nu^\psi \mid J/\psi>
<K J/\psi \mid \tilde{O}_2 \mid B><B \mid j^B \mid 0>}{(m_K^2-q^2)
(m_{\psi}^2-p^2)(m_B^2-P^2)} ~.
\label{phys}
\ee
However, this problem is much more difficult than it sounds. One
reason is that apart from the ground state contribution
(\ref{phys}) and analogous contributions from excited states, one
also has to take into account contributions to the spectral
functions from continuum states with flavour quantum numbers which
differ from those of the respective currents.
Such intermediate states emerge in every channel of the correlation
function due to final state interaction. For
example, in the $B$-meson channel the following contribution
appears inevitably:
\be
\tilde{\Pi}_{\mu\nu}^{"DD^*_s"} =
i\frac{<0 \mid j_{\mu5}^K \mid K><0\mid j_\nu^\psi \mid J/\psi>
<K J/\psi \mid "D D^*_s"><"DD^*_s"\mid \tilde{O}_2~j^B \mid 0>}
{(m_K^2-q^2)(m_{\psi}^2-p^2)(m_{DD^*_s}^2-P^2)}
\label{paras}
\ee
Here, the questionable intermediate state carries the quantum
numbers of a virtual $DD_s^*$ state. It is created by weak
interaction and converted into the $J/\psi K $ final state by
strong interaction.
The problem is that this continuum of states has a mass-threshold
below the $B$-meson pole. It can
therefore not be suppressed by Borel transformation and subtracted
away similarly as outlined in section 2 for the normal continuum
in the $B$-channel starting at a threshold $s_0 > m_b^2$. On the
contrary, this unwanted contributions will be enhanced by Borel
transformation relative to the contribution from the ground state
$B$ roughly by a factor $exp[(m_b^2-4m_c^2)/M^2]$ which is quite
large at characteristic values of the Borel parameter
$M^2 \simeq (m_B^2-m_b^2)$.

Such dangerous contributions are also present in D decays. They
have been estimated in ref. [11] using a simple model and found to
be unimportant. Unfortunately, in the present case the solution of
this problem seems to be less trivial. We have identified the
analogous contributions in the QCD diagrams of Fig. 4 and find that
they are numerically very important after Borel transformation.
Indeed, examination of the analytical properties of the diagrams in
Fig. 4 shows that in addition to a pole at $P^2=m_b^2$
the four quark condensate diagram, Fig. 4c, has a discontinuity
at $P^2 \geq 4m_c^2$ connected with the four-quark
intermediate state with the flavour combination
$\overline{u}s\overline{c}c$.
This is just the flavour composition of the virtual
$DD^*_s$ continuum in (\ref{paras}). It seems to be reasonable to
again use the quark-hadron duality principle in order to cancel
this piece of (\ref{res}) against the unwanted hadronic
contribution (\ref{paras}). Practically, this
results in neglecting the four-quark condensate diagram
altogether.

{}From here on, we follow the usual procedure. We perform a
Borel transformation in $P^2$ and take
moments in $p^2$ as is usually done in the charmonium channel.
The momentum squared in the $K$-meson channel is
kept fixed at $Q^2=-q^2 \simeq 1GeV^2$. In order to
take into account the higher states in the resulting
sum rules we adopt for simplicity a two-resonance description in
each of three channels,  i.e. above the effective thresholds
$s_{0i}$ we describe the hadronic spectral functions
by one effective resonance with a mass equal to $\sqrt{s_{0i}}$.
Normalizing the matrix element (\ref{nf}) so that the amplitude
(\ref{matr}) is proportional to combination
\be
a_2f_K^+=(c_2+\frac{c_1}3)f_K^+ + 2c_1\tilde{f}~,
\label{final}
\ee
we obtain for the nonfactorizable contribution
\be
\tilde{f} = -(0.045-0.075)~.
\label{ftilde}
\ee
The range reflects our estimate of the uncertainty in the QCD
calculation and the continuum subtraction.

Our analysis shows that the factorizable, nonleading in $1/N_c$
term and the nonfactorizable term in (\ref{final})
are opposite in sign. Quantitatively, the nonfactorizable matrix
element is considerably smaller than the factorizable one,
$|\tilde{f}/f^+_K(m_\psi^2)|\simeq O(10 \%)$. Nevertheless, the
nonfactorizable contribution to the decay amplitude is very
important because of the large coefficient,
$2c_1/(c_2+c_1/3) \simeq 20 \div 30 $. In fact, if
$|\tilde{f}|$ is close to the upper end of the range given in
(\ref{ftilde}), the nonfactorizable contribution almost cancels
the factorizable one proportional to $c_1/3$, thus leading to the
prediction (\ref{BRNc}) which agrees with experiment. This is
exactly the scenario anticipated by employing the $1/N_c$-rule.

At the same time, in our approach the nonfactorizable
contributions are  expected to be channel-dependent on quite
general grounds. This expectation is corroborated by our
preliminary analysis of other nonleptonic $B$-decays. In other
words, as is obvious from eq. (\ref{final}) there is no
theoretical reason to expect a single universal value for the
effective coefficient $a_2$.
Even the sign of $a_2$, depending on the sign of $\tilde{f}$
in a given channel, may vary. Universality can
at most be anticipated for certain classes of decay modes, such as
$B\rightarrow D\pi$ or $B \rightarrow D\overline{D}$, etc.
Moreover, there is no simple relation between $B$ and $D$ decays
in our approach since the  OPE for the corresponding correlation
functions significantly differ in the relevant diagrams and in the
hierarchy of mass scales.
We hope to be able to clarify this issue further.
Theory seems to predict a much richer pattern in two-body weak
decays than what is revealed by the present phenomenological
approach to the data.

\section{Conclusion}

We have discussed applications of QCD sum rules in different
variants to form factors and exclusive decay amplitudes of $B$
mesons, both being  important issues in $B$-physics.
There are many other interesting aspects which can be studied in
the framework of QCD sum rules and which have not been touched upon
here. We just mention the dependence of form factors and couplings
on the heavy quark mass. Here, the sum rule technique can be employed
complementarily to the heavy-quark effective theory.
{}From a
theoretical point of view, heavy-light bound
states such as the $B$ meson are very suitable systems for
studying the dynamics of the light quark and gluon degrees of
freedom and for probing nonperturbative methods in QCD.
It will be interesting to see whether or not the QCD sum rule
technique passes these tests as successfully as the various tests
in the past. For the time being, it is the only viable approach to
some problems such as the calculation of weak decay amplitudes
beyond factorization. Even the capabilities of current lattice
calculations do not allow to solve this problem
(as remarked in ref. [49]).

\section{Acknowledgements}
We are grateful to V. Belyaev and B. Lampe for collaboration
and discussions. Without their contributions this report would not
have been written. We also acknowledge helpful discussions with
M. Shifman. A. Khodjamirian thanks the TPI , University of Minnesota
for supporting his participation in the Workshop.

\newpage
\begin{center}
{\bf Figure Captions}
\end{center}
{\bf Fig. 1}: QCD diagrams contributing to the matrix element
(2) involving (a) quark-antiquark light-cone wave functions; (b)
three-particle quark-antiquark-gluon wave functions;
(c) and (d) perturbative $O(\alpha_s)$ corrections.
Solid lines represent quarks, dashed lines gluons,
wavy lines are external currents.

\vspace{0.5cm}

{\bf Fig. 2}: $B$-meson form factors calculated from
light-cone sum rules: (a) the form factor $f^+_{\pi}(p^2)$ of
$B \rightarrow \pi $ transitions and (b) the form factor
$f^+_{K}(p^2)$ of $B \rightarrow K $ transitions
at $M^2=10GeV^2$ (upper solid curves) and  $M^2=15GeV^2$
(lower solid curves). The quark
model predictions from ref. [38]
(dash-dotted curves) and the QCD sum rule results for
$f^+_{\pi}(p^2)$ from ref. [20] (dashed curve) and for
$f^+_{\pi}(0)$ from ref. [6] (arrow) are shown for
comparison.

\vspace{0.5cm}

{\bf Fig. 3}: Diagrams associated with the correlation function (24):
(a) bare diagram with $H_W$ replaced by $O_2$; (b) and (c)
diagrams corresponding to perturbative gluon
corrections.

\vspace{0.5cm}

{\bf Fig. 4}: Diagrams associated with (a) the gluon condensate,
(b) the quark-gluon condensate and (c) the four-quark condensate
contributions to the correlation function (24) with $H_W$
replaced by $\tilde{O}_2$ .

\end{document}